\documentclass[%
    reprint,
    nofootinbib,
    amsmath,
    amssymb,
    aps,
    prab,
    longbibliography,
    floatfix,
]{revtex4-2}

\usepackage{graphicx}
\usepackage{dcolumn}
\usepackage{bm}
\usepackage{booktabs}
\usepackage{physics}
\usepackage{siunitx}
\usepackage[justification=RaggedRight]{caption}
\usepackage{subcaption}
\usepackage{multirow}
\usepackage{hyperref}

\begin{document}

\title{Transverse beam instabilities in low-emittance booster synchrotrons}

\author{W. Foosang}

%\email{watanyu.foosang@synchrotron-soleil.fr}
\author{A. Gamelin}
\email{alexis.gamelin@synchrotron-soleil.fr}
\author{V. Gubaidulin}
\author{R. Nagaoka}
\affiliation{Synchrotron SOLEIL, L'Orme des Merisiers, 91190, Saint-Aubin, France
}

\begin{abstract}
As the ring-based light source community is moving towards fourth-generation light sources, many facilities plan to upgrade their boosters in parallel to meet the more demanding beam properties for the storage ring, especially in terms of a much lower emittance. Concerns over collective effects have, therefore, risen, particularly in the transverse planes, since the vacuum chamber dimensions tend to be reduced as a way to achieve a stronger focusing force on the beam. In this article, we present numerical studies on transverse beam instabilities, both in the single- and multibunch regimes, in the SOLEIL\,II booster as an example of a low-emittance booster. We show that Landau damping is an efficient mechanism for suppressing both transverse single-bunch and coupled-bunch instabilities. We also prove that the damping in the longitudinal plane can diffuse to the transverse plane and limit the transverse emittance growth. Moreover, we have discovered that the beam can exhibit sawtooth instability at high energy and that broad-band impedance is one of the key factors in suppressing transverse coupled-bunch instability.
\end{abstract}

\maketitle

\section{Introduction}\label{sec:intro}
Booster synchrotrons generally refer to ring-based accelerators used for ramping the energy of a particle beam coming from a low-energy section before injecting it into a higher-energy ring in the accelerator complex. The conventional acceleration scheme in many light sources begins with a linear accelerator (linac) where the energy is escalated from the order of keV to MeV, then with a booster to GeV, and finally, a storage ring where the energy remains constant at the GeV scale. Being an intermediate accelerator through which the beam passes in less than a second, collective effects in light source's boosters are often not profoundly studied.

However, this may no longer be acceptable for the new generation boosters built for 4\textsuperscript{th} Generation Light Sources (4GLSs) since they need to provide a much lower emittance beam at extraction to match the usually small dynamic aperture of 4GLS storage rings. Taking for example SOLEIL\,II \cite{CDR:SOLEIL, TDR:SOLEIL, susini:EPJ2024}, the upgrade project towards a 4GLS of SOLEIL, the horizontal beam emittance at the new booster's extraction will be \SI{5.2}{\nano\meter\radian}, compared to \SI{140}{\nano\meter\radian} in the existing booster. This new value is almost identical to the natural emittance in the present 3GLS storage ring of SOLEIL. Such a low emittance value of the new booster imposes that the beam cannot tolerate instabilities that will explode the emittance as much as it could have for 3GLSs. Thorough collective effects studies in a 4GLS booster, therefore, need to be carried out.

In Ref.\,\cite{XU:2019313}, where Transverse Single-Bunch Instabilities (TSBI) were studied in the HEP booster, it was shown that including an energy ramp and nonlinearities from sextupoles during particle tracking can increase the instability threshold. Then, recent studies from Diamond-II \cite{husain:fls2023-tu4p19, husain:ipac2024-tupg18} were extended to Transverse Coupled-Bunch Instabilities (TCBI) and included a physical aperture in the tracking with energy ramp. It was concluded that, despite the inclusion of the lattice nonlinearities, some charge can be lost between injection and extraction. A transverse feedback system is therefore envisaged in order to increase the charge threshold in the Diamond-II booster. These results reflect the importance of collective effects in the low-emittance boosters.

This article puts together studies on TSBI and TCBI in a low-emittance booster synchrotron that will serve as an injector for a 4GLS storage ring. We present an observation of a sawtooth instability in the single-bunch regime during the booster ramp process, as well as a demonstration that longitudinal synchrotron radiation damping can also give a damping effect in the transverse planes. We also prove that the increase of instability threshold under the presence of nonlinearities is due to the Landau damping effect, which can be effective against both TSBI and TCBI. Lastly, we show that broad-band impedance can result in a suppression of TCBI at low chromaticity. While the effect of these elements might be well known individually in storage rings, we present a comprehensive study in the context of a booster machine. This work demonstrates the relevance of these effects to the design of future ultra-low emittance booster synchrotrons.

The SOLEIL\,II booster is used as an example for this study. It has a two-fold symmetry 16-Bend Achromat (16BA) lattice \cite{tordeux:ipac2021-mopab248, tordeux:ipac2021-mopab113, tordeux:ipac2023-wepl072} designed to fit in the existing booster tunnel. Its main parameters can be found in Table\,\ref{tab:intro:beam_param_booster_upgrade}. Note that the emittance, bunch length, and energy spread in this table are the equilibrium values at the corresponding energy. As the beam can never reach the equilibrium at 150\,MeV during the ramp, the values appear in the 150\,MeV column in this table, therefore, do not show the beam parameters at injection. These are instead shown separately in Table\,\ref{tab:intro:beam_param_injection} for clarity. Two different bunch lengths at injection are used depending on the regime being studied. This is in accordance with the two operation modes of the upstream linear accelerator: the short-pulse mode for single-bunch, and the long-pulse mode for multibunch operation mode of the storage ring \cite{foosang:tel-04496043}. The injected beam also has an equal emittance in the horizontal and vertical planes. The booster impedance model can be consulted in \cite{foosang:tel-04496043}. This study can be of great relevance to other low-emittance boosters, especially those foreseen with a high-charge operation mode for a swap-out injection scheme or with a small vacuum chamber radius since these configurations can give an immense rise to the machine coupling impedance, hence possibly strong collective effects.

The main tool used in this work is the in-house developed tracking code \texttt{mbtrack2} \cite{mbtrack2:ipac2021-mopab070, zenodo_mbtrack2} capable of simultaneous single-bunch and multibunch collective effects calculations. The studies are done both at a fixed energy and with an energy ramp. The latter is done to imitate the booster ramping process during which the equilibrium parameters vary continuously. This technique allows us to clearly see the beam evolution along the ramp and increases the accuracy of the results compared to studying at a fixed energy alone. We also include the nonlinearity from sextupoles, which yields the Amplitude-Dependent Tune Shift (ADTS) and can induce the Landau damping effect as it was shown in the storage ring of ELETTRA \cite{Tosi:PhysRevSTAB.6.054401}. It can play an important role in beam dynamics since a 4GLS booster tends to have much stronger sextupoles to correct the highly negative natural chromaticity caused by strong quadrupoles. The ADTS coefficients of the SOLEIL\,II booster used in this work can be found in Appendix \ref{app:adts}.

This article is separated into two main parts: Section \ref{sec:SB} presents the study in the transverse single-bunch regime, and Section \ref{sec:MB} presents the multibunch one. In the first part, we show the use of a stability diagram to assess the beam stability at injection, followed by simulations with an energy ramp that can reveal emittance growth along the ramp, if it exists. We also present in this part the observation of sawtooth instability and the effect of longitudinal damping on TSBI. In the second part, a general trend of the TCBI along the ramp in a booster is given via a simple mathematical model. Then, tracking simulation results are presented and prove a certain level of accuracy of this model. In this part, we will see that TCBI is mitigated not only by ADTS but also by broad-band impedance, where the latter is generally unwanted because it can excite single-bunch instability.

\begin{table}[htb]
   \centering
   \caption{SOLEIL\,II booster equilibrium parameters (H, V, and L denote the horizontal, vertical, and longitudinal planes, respectively.)}
   \begin{ruledtabular}
   \begin{tabular}{lcc}
       Parameter & \multicolumn{2}{c}{Value at} \\
       & 150 MeV & 2.75 GeV  \\
       \midrule
           Circumference                & \multicolumn{2}{c}{156.46 m}          \\
           Repetition rate              & \multicolumn{2}{c}{3 Hz}              \\
           Betatron tune H/V            & \multicolumn{2}{c}{13.19/4.19 }       \\
           Momentum comp. factor        & \multicolumn{2}{c}{\SI{3.3e-3}{} }    \\
           Nat. chromaticity H/V        & \multicolumn{2}{c}{-26/-13 }          \\
           RF voltage                   & 164 kV   & 3 MV                      \\
           Natural emittance            & 15.3 pm rad   & 5.2 nm rad            \\
           Natural bunch length         & 0.3 ps        & 25 ps                 \\
           Natural energy spread        & \SI{5.1e-5}{} & \SI{9.3e-4}{}         \\
           Damping time H/V/L           & 20/31/23 s    & 3.3/5.2/3.7 ms        \\
           Energy loss per turn         & 4.9 eV        & 554 keV               \\
   \end{tabular}
   \end{ruledtabular}
   \label{tab:intro:beam_param_booster_upgrade}
\end{table}

\begin{table}[htb]
   \centering
   \caption{Input beam parameters at injection used in this study}
   \begin{ruledtabular}
   \begin{tabular}{lccc}
       Parameter & Unit & \multicolumn{2}{c}{Value} \\
                 &      & single-bunch & multibunch \\
       \midrule
        4 $\times$ rms H\&V emittance  & \SI{}{\micro\meter\radian}    &  \multicolumn{2}{c}{0.68 (round beam)}\\
        Rms energy spread   & -                             &  \multicolumn{2}{c}{\SI{2.83e-3}{}}   \\
        Rms bunch length    & ps                            &  200  &   150                         \\
        Nominal charge      & nC                            &  0.5  &   3                           \\   
        Maximum charge      & nC                            &  1.5  &   10                          \\
   \end{tabular}
   \end{ruledtabular}
   \label{tab:intro:beam_param_injection}
\end{table}

\section{Transverse single-bunch instabilities} \label{sec:SB}
\subsection{At injection energy}
We commence the study in the single-bunch regime and we will consider only the injection energy first. This is because collective effects are usually strong at low energy and they become weaker at higher energy. If no instability can be observed at this point, a stable beam can usually be inferred for the rest of the ramp. 

One of the most general mitigation mechanisms for most collective instabilities is synchrotron radiation, but it is very weak at the energy scale of hundreds of MeV, as we will see in Section \ref{sec:general_trend_TCBI}. However, as a low-emittance booster will likely have strong sextupole magnets, their strong nonlinear fields will lead to a highly nonlinear tune shift, i.e., strong ADTS. The beam will then have a large incoherent tune spread (rms tune spread $\approx$ \SIrange{5e-4}{e-3}{}) as a consequence, leading to suppression or weakening of beam instabilities via Landau damping. The result can be that the instability totally vanishes or its magnitude is attenuated. This mechanism is known as Landau damping \cite{Landau_collection, Hereward:1542436, Herr:Intro_to_landau} and it has always been a common mitigation method for hadron machines \cite{metral:IEEE-2016, Metral:EPJP2021, Buffat:PhysRevSTAB2014} since synchrotron radiation is usually too weak for such particles.

We will, therefore, compare instability growth rates to the limit of the Landau damping to assess the stability at the injection energy. To do so, one can draw a \textit{stability diagram} which maps between the complex tune shift when ADTS is present and that in the case where ADTS is absent. In this article, we will denote the first quantity as $\Delta Q_{\text{coh}}$ and the second as $\Delta Q_{\text{coh}}^{\text{lin}}$ to highlight the fact that the latter is obtained in the absence of ADTS, i.e., without the nonlinearity. To construct a stability diagram, we need to extract $\Delta Q_{\text{coh}}^{\text{lin}}$ from simulations and plot the imaginary part against their real part. Physically interpreted, the former represents the instability growth rate, while the latter represents the coherent tune shift, the Fourier transform of the center-of-mass position of the bunch. Then, on the same space, we overlay a curve called the \textit{stability limit} which marks where zero growth rate will be if ADTS is present. If the $\Delta Q_{\text{coh}}^{\text{lin}}$ data point falls under the curve, it means that the beam will be stable under ADTS. In contrast, if the data point is above the stability limit, it means that the beam will be unstable even if the ADTS is present.

Strictly speaking, the stability limit is the analytical solution to the dispersion relation \cite{jsberg:landau-damping, burov_nested_2014} where the growth rate of a bunched beam under ADTS is zero; $\Im[\Delta Q_\text{coh}]= 0$. This stability boundary should agree with the tracking simulation results; see, e.g.,~\cite {gubaidulin_landau_2022}.

% The diagram is composed of the imaginary and the real part of the complex coherent tune shift, obtained in the absence of ADTS.

% The diagram is composed of the imaginary and the real part of the complex coherent tune shift in the absence of ADTS, that is, without nonlinearities. This quantity will be denoted by $\Delta Q_{\text{coh}}^{\text{lin}}$. Then, 

% space between the imaginary part of the complex coherent tune shift $\Im[\Delta Q_\text{coh}]$ and its real part $\Re[\Delta Q_\text{coh}]$. Physically interpreted, the former represents an instability growth rate, while the latter represents the coherent tune shift, the Fourier transform of the center-of-mass position of the bunch. It should be precised that the tune shifts 

% The curve in the diagram shows the analytical solution to the dispersion relation \cite{jsberg:landau-damping, burov_nested_2014} where $\Im[\Delta Q_\text{coh}]= 0$ in the case of a bunched beam with ADTS. This stability boundary should agree with the tracking simulation results; see, e.g.,~\cite {gubaidulin_landau_2022}.

% To determine the beam stability, we must track the beam without ADTS to observe the intrinsic properties without the intervention of Landau damping, then extract the coherent tune shift and the instability growth rate. If the data point falls in the area under the stability limit curve, it means that the instability can be stabilized since the growth rate is less than the Landau damping rate. Otherwise, if the data point falls outside this area, the beam will be unstable.

\begin{figure}[tb]
    \centering
    \includegraphics[width=\linewidth]{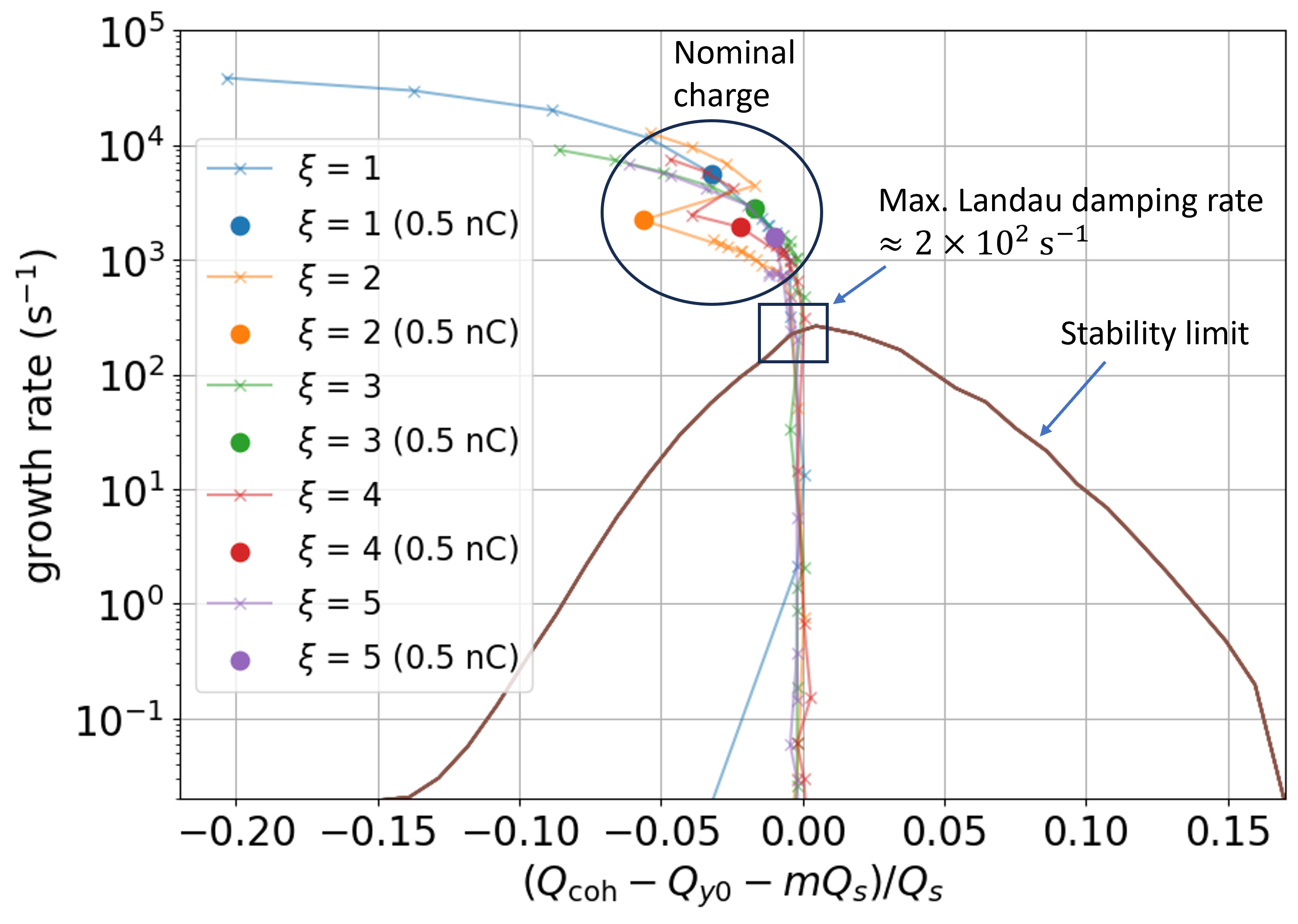}
    \caption{Stability diagram of the SOLEIL\,II booster at the injection energy. The markers show the head-tail instability growth rate at different chromaticities $\xi$ without ADTS, while the curve shows the stability limit solved analytically. To highlight the results at the nominal charge, they are shown in circle markers.}
    \label{fig:SB_chargescan_inj}
\end{figure}

Figure \ref{fig:SB_chargescan_inj} shows the stability diagram of the SOLEIL\,II booster at 150\,MeV, where we can see the stability limit and transverse $\Delta Q_{\text{coh}}^{\text{lin}}$ extracted from tracking simulations with different bunch charges. The horizontal axis is the mode tune shift where $Q_\text{coh}$ is the coherent tune obtained from the simulation, $Q_{y0}$ is the zero current coherent tune in the vertical plane, $m$ is the head-tail mode number, and $Q_s$ is the synchrotron tune. This is a standard way to visualize the data, instead of showing the raw coherent tune shift, it shows the detuning from the corresponding head-tail mode in the unit of $Q_s$. In this plot, the charge is varied between \SIrange{0.001}{1.5}{\nano \coulomb}. The data point for each chromaticity $\xi$ \footnote{The chromaticities throughout this article are non-normalized chromaticities.} evolves from the lower center of the diagram at a low charge, then climbs up vertically as the charge increases until it reaches the stability limit. It finishes in the upper left corner area at high charge at last. We can see that, at the nominal charge of 0.5\,nC, the beam finds itself lying outside the stability limit with a growth rate in the order of \SI{e3}{\per\second} while the maximum Landau damping rate is about \SI{2e2}{\per\second}. 

According to this result, the beam stability cannot be assumed for the rest of the ramp. However, as the head-tail growth rate decreases with energy, while the synchrotron radiation damping rate increases, it is possible that the instability will not appear at all. Therefore, simulations with the energy ramp are needed to evaluate the instability along the ramp.

\subsection{Along the ramp}
We perform tracking simulations with \texttt{mbtrack2}, which uses the one-turn maps tracking method. A variation of beam energy is added to predict the realistic evolution of the beam throughout the ramp. The principle is to update, at each turn, the beam energy and equilibrium parameters such as energy loss per turn, radiation damping time, rf voltage and synchronous phase, natural energy spread, natural emittance, etc.
Longitudinal short-range wakefields are always included in the simulations.
Radiation damping time is included in all the simulations, unless we explicitly write that it has been excluded in one of the planes.%\footnote{Turning off longitudinal radiation damping time is not possible because it will lead to the loss of beam charge during the ramp.}.
The beam energy variation itself is calculated via the energy gain per turn equation in relation to the dipole field ramping rate, and the equilibrium parameters are calculated using their proportionality to the beam energy. More details on the booster ramp model can be consulted in \cite{Foosang_2024, foosang:tel-04496043}.
Moreover, the beam injected into the booster is not matched to the booster equilibrium parameters at the injection energy. Parameters of the injected beam, determined by the linac and the transfer line, are shown in Table~\ref{tab:intro:beam_param_injection}.

\begin{figure}[tb]
    \centering
    \begin{subfigure}[b]{0.49\textwidth}
        \includegraphics[width=1\textwidth]{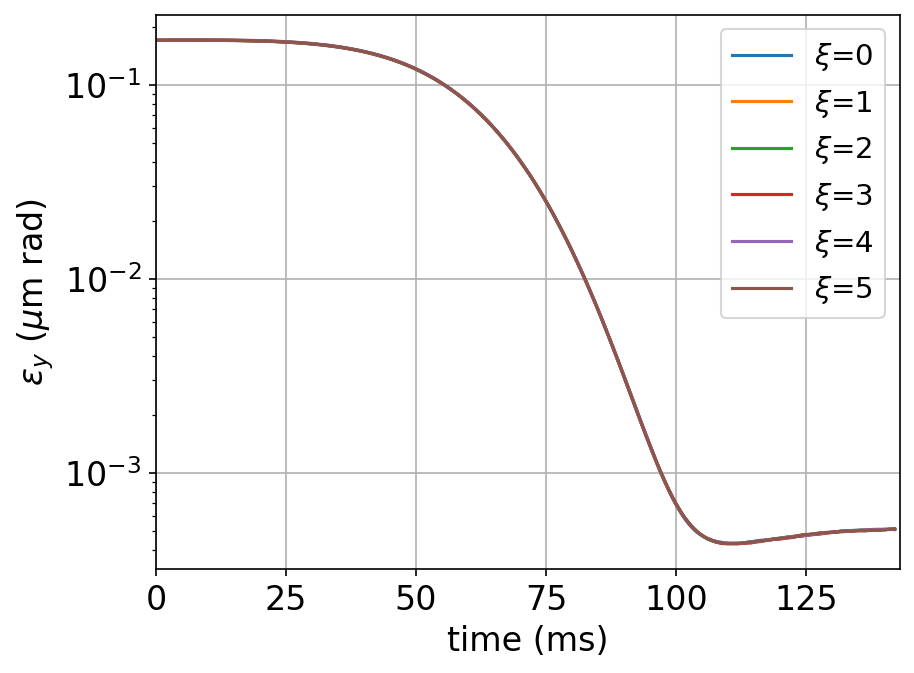}
        \caption{q = 0.5 nC}
        \label{fig:SB_ramp_0_5nC}
    \end{subfigure}
    \begin{subfigure}[b]{0.49\textwidth}
        \includegraphics[width=1\textwidth]{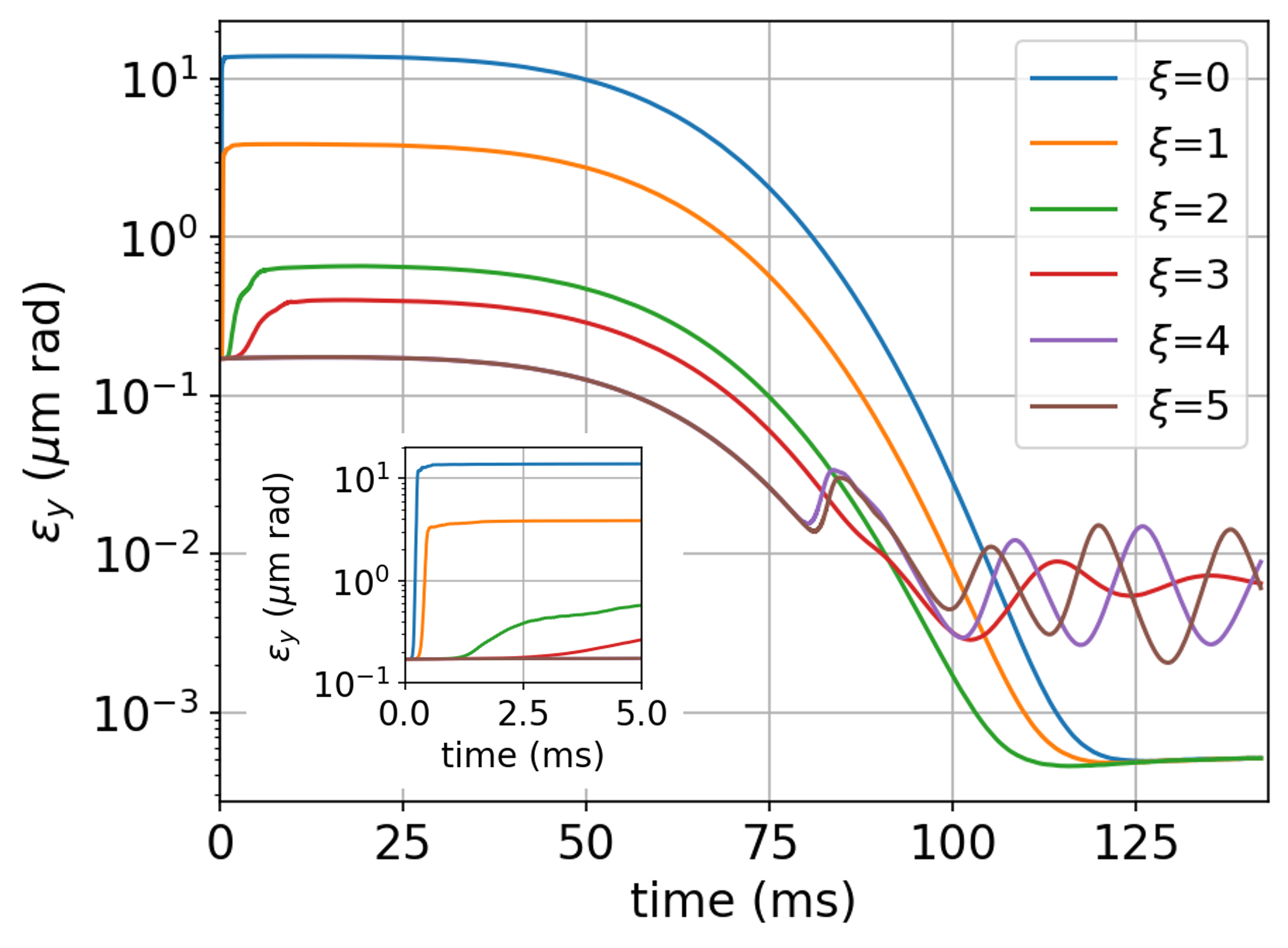}
        \caption{q = 1.5 nC}
        \label{fig:SB_ramp_1_5nC}
    \end{subfigure}
    \caption{Vertical single-bunch emittance along the ramp at different chromaticities at (a) the nominal charge, and (b) the maximum charge.}
    \label{fig:SB_ramp}
\end{figure}

The nominal and maximum single-bunch charges of the SOLEIL\,II boosters, 0.5 and 1.5\,nC, are chosen as our strategic points for simulations along the ramp to obtain results for the nominal operation mode and the margin. The vertical beam emittance as a function of time after injection at different chromaticities is presented in Fig.~\ref{fig:SB_ramp}. Note that, in \texttt{mbtrack2}, the emittance in a given transverse plane is computed from $\sqrt{\expval{u_i^2}\expval{u_i'^2}-\expval{u_iu_i'}^2}$ where the angle bracket means average value, and $u_i$ and $u_i'$ are the betatron oscillation amplitude of a single particle and its derivative, respectively. Through this work, we used $\SI{1e5}{}$ macro-particles per bunch and the averaged beta-functions $(\beta_x,\beta_y)=(\SI{5.69}{\meter},\SI{7.72}{\meter})$ with $x$ and $y$ respectively indicating the horizontal and vertical plane.

\begin{figure}[tb]
    \centering
    \includegraphics[width=1\linewidth]{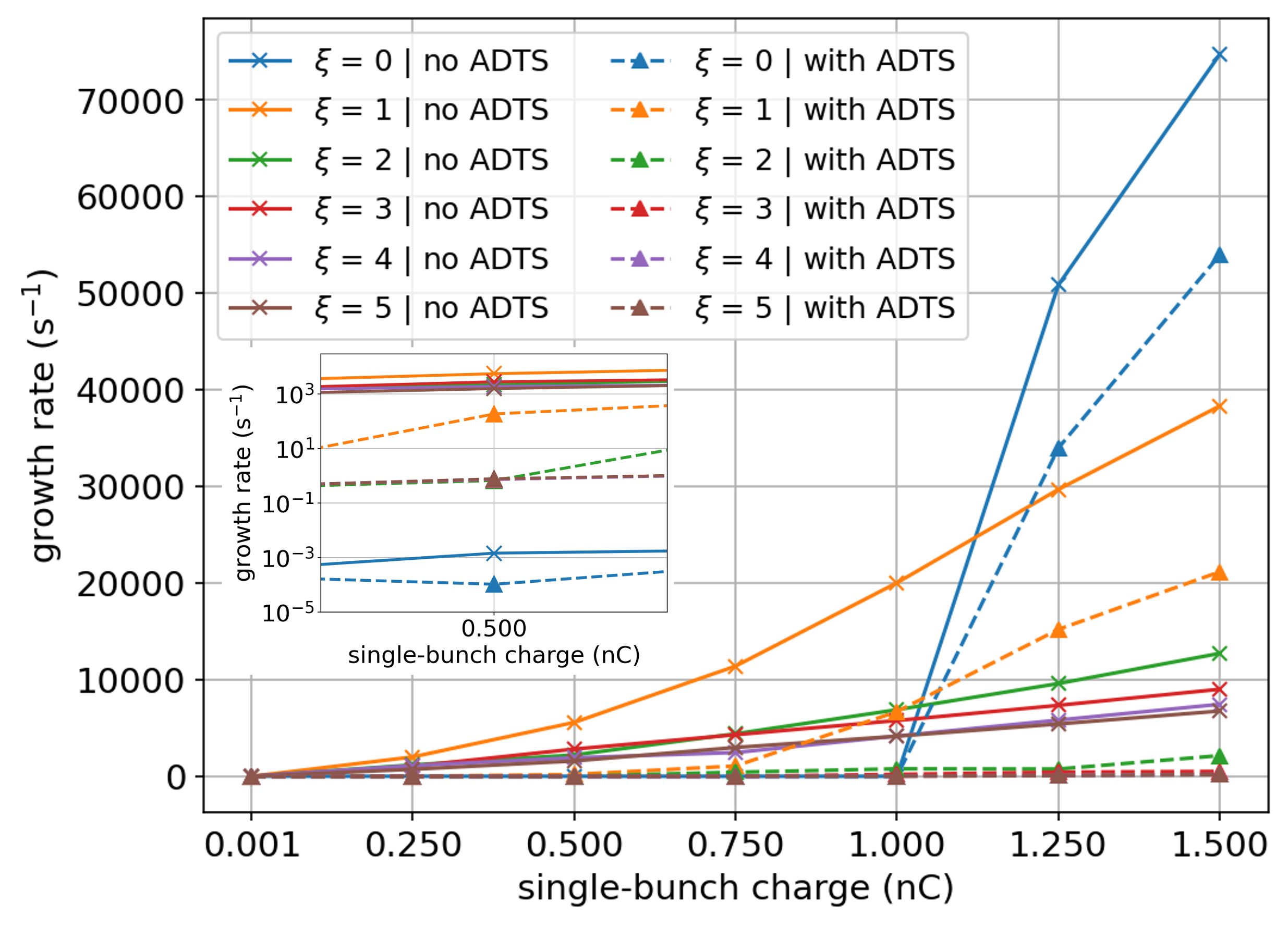}
    \caption{Comparison of instability growth rates at 150\,MeV with and without ADTS. The small figure zooms in around 0.5\,nC with the vertical axis in log scale.}
    \label{fig:SB_all_growth_rate_inj}
\end{figure}

For the nominal charge in Fig. \ref{fig:SB_ramp_0_5nC}, it turns out that the beam is well stable and no sign of instability can be observed regardless of chromaticity. The vertical emittance is damped all the way from injection and reaches the designed value 0.5\,nm\,rad at the extraction, which is the desirable situation. This result is rather surprising since some emittance increase could have been expected due to the high growth rate above the Landau stability limit shown in Fig.\,\ref{fig:SB_chargescan_inj}. The explanation for this is that even though Landau damping cannot totally suppress the instability, the growth rates are significantly reduced by the presence of ADTS, in addition to the reduction due to the increasing beam energy. This can be seen in Fig.\,\ref{fig:SB_all_growth_rate_inj} where growth rates extracted from the tracking simulations with and without ADTS are compared. For a certain chromaticity, the ADTS can greatly diminish the instability growth rate even at zero chromaticity where the Transverse Mode-Coupling Instability (TMCI) \cite{Kohaupt:IEEE1979, Ruth:IEEE1983, Chao:1993zn}, which is extremely strong once emerged, takes place. At 0.5\,nC for example, the growth rates can drop from \SI{e3}{\per\second} to the minimum around \SI{e-1}{\per\second} which is less than or in the same order as the booster's half-cycle rate \SI{6}{\per\second}. This means that the instability will never have enough time to build up before reaching the extraction, and the beam is simply damped by synchrotron radiation as the energy increases. 

\begin{figure}
    \centering
    \includegraphics[width=1\linewidth]{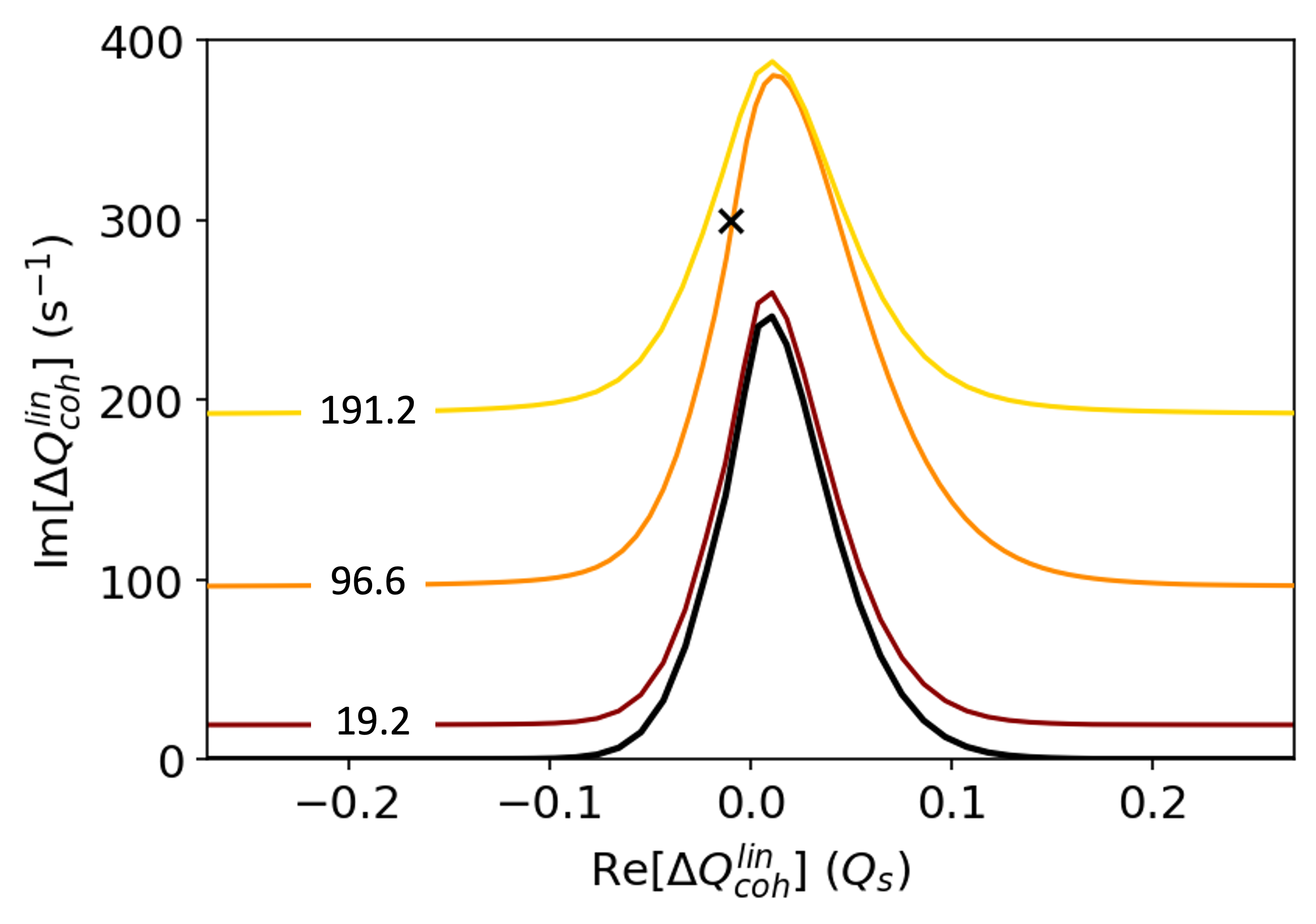}
    \caption{Stability limit (in black) and isolines of instability growth rate (in colors).}
    \label{fig:stability_diagram_varied_epsilons}
\end{figure}

The growth rate reduction outside the stability limit can actually be viewed as a distortion of the complex tune shift space due to the presence of nonlinearity as we can see in Fig.\,\ref{fig:stability_diagram_varied_epsilons}. Here, the same stability limit shown previously is drawn in black. However, this is only a curve that corresponds to Im[$\Delta Q_\text{coh}$]=0, the definition of a stability boundary. It means that the growth rate (under ADTS) along this curve is zero, and we can also call it an \textit{isoline} of the zero growth rate. There are, in fact, an infinite number of isolines that show other growth rates Im[$\Delta Q_\text{coh}$] (see e.g \cite{Schenk:PhysRevAccelBeams.21.084402}). Here in Fig. \ref{fig:stability_diagram_varied_epsilons}, we plot some others showing Im[$\Delta Q_\text{coh}$]=19.2, 96.6 and 191.2\,\SI{}{\per\second}. It can be clearly seen that the neighboring space of the stability limit is distorted. If we have a $\Delta Q_{\text{coh}}^{\text{lin}}$ falling, for example, at the point marked as ``x", it means that the growth rate of \SI{300}{\per\second} without ADTS will be reduced to about \SI{100}{\per\second} under the ADTS.

Moving back to the maximum charge case in Fig. \ref{fig:SB_ramp_1_5nC}, the situation gets more complicated than in the nominal charge one. An emittance blow-up can be clearly seen right after the injection for chromaticities 0 to 3. It is, however, limited by the ADTS once the emittance is large enough so that the ADTS is sufficiently strong to counteract the instability. But, as one can expect from any countermeasure, the higher the growth rate, the less efficient the ADTS. This is the reason why the emittance at low chromaticities, whose growth rate is large, can reach a higher value than the one at higher chromaticities with smaller growth rates.

\subsubsection{Sawtooth instability}
Unique to the case of maximum charge, an unexpected sawtooth feature was observed at chromaticities 3 to 5. Further investigations were, therefore, carried out by simulating TSBI at different energies and the growth rates were compared to the analytical synchrotron radiation damping rate. It should be noted that, in this case, synchrotron radiation was not included in the transverse planes during the tracking in order to speed up the calculation and because we were interested in the pure instability growth rate. 
It was, however, included in the longitudinal plane to retain all the particles within the rf bucket\footnote{The rf bucket shrinks during the ramp, it is not an issue normally as the beam longitudinal emittance is also decreasing during the ramp due to radiation damping. But if the longitudinal synchrotron radiation damping is switched off, some particles will escape the rf bucket during the ramp, which disturbs the longitudinal binning profile used to compute longitudinal and transverse wake potentials.}.
%Because otherwise, a large fraction of the particles will be lost since we simulate the beam with injection parameters (see Section~\ref{tab:intro:beam_param_injection}). 

\begin{figure}[tb]
    \centering
    \includegraphics[width=1\linewidth]{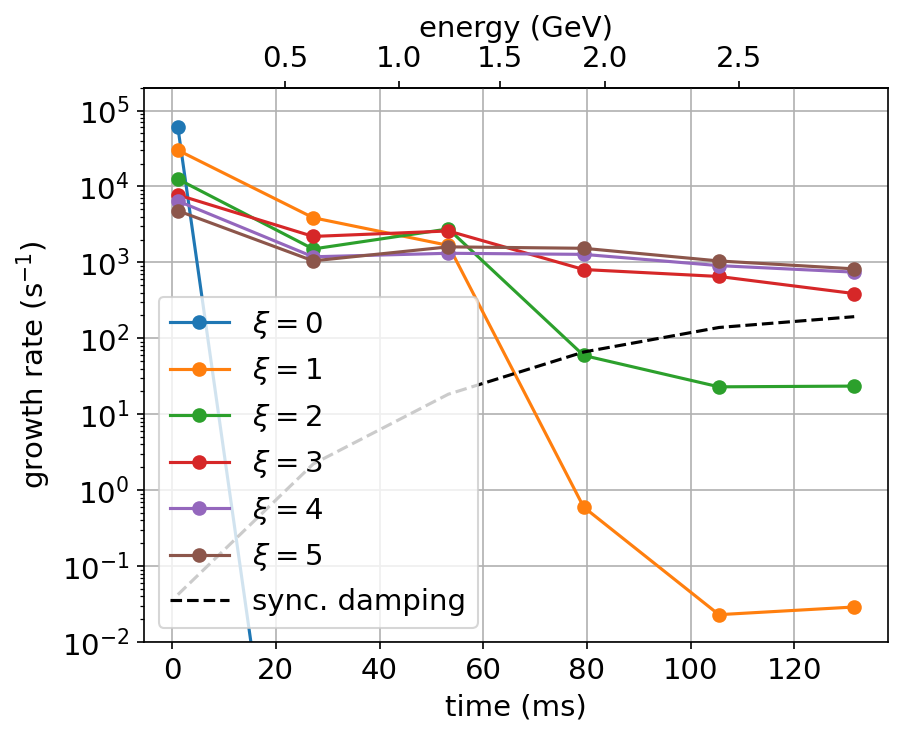}
    \caption{TSBI growth rate without ADTS at 1.5\,nC at different points on the ramp compared to synchrotron radiation damping rate.}
    \label{fig:gr_scan_mbtrack2}
\end{figure}

The investigation results are shown in Fig. \ref{fig:gr_scan_mbtrack2}. Here, the growth rate of a 1.5\,nC bunch at different chromaticities is determined at various points along the ramp discretely, meaning that each point represents an independent simulation with a fixed beam energy. We can see that the growth rates of the high chromaticity group (3 to 5) always remain above the radiation damping rate, while those of the low chromaticity group (0 to 2) drop below the latter at some point. Note that the growth rate at zero chromaticity has dropped below \num{e-2} since the first 20\,ms. As we have seen in Fig. \ref{fig:SB_ramp_1_5nC} that only in the high chromaticity group did the sawtooth behavior occur, it is clear that the growth rate being stronger than radiation damping is the reason for the regrowing emittance. Another factor that supports this regrowth is the ineffectiveness of the ADTS at this point of the ramp due to the small emittance. However, as the emittance increases from the instability, the effect of ADTS becomes stronger and can damp the beam again. This mechanism repeats itself as long as the growth rate is larger than the radiation damping rate and it is the explanation to the sawtooth-like shape of the emittance. 

Such an instability was also observed in the storage ring of APS in 1999 \cite{Harkay:1999} and Landau damping was used to explain the self-limiting character \cite{Harkay:2001}. More recently, MAX\,IV published a similar observation when the influence of ADTS on TMCI was being studied \cite{Brosi:PRAB2024}.

\subsubsection{Effect from the longitudinal damping}
Another interesting point in Fig.\,\ref{fig:gr_scan_mbtrack2} is around 80\,ms where the growth rate at chromaticity 1 and 2 drops below the synchrotron damping rate and it appears to coincide with the point where the sawtooth instability emerges in Fig.\,\ref{fig:SB_ramp_1_5nC}. The question arose as to what caused the abrupt fall in these two mentioned cases when there was no synchrotron radiation in the transverse planes nor ADTS to damp the beam. To answer the question, we looked into the only damping mechanism present in the simulations for Fig.\,\ref{fig:gr_scan_mbtrack2}, the synchrotron radiation in the longitudinal plane. We repeated the simulation at 80\,ms, corresponding to 1.9\,GeV, and varied the longitudinal synchrotron damping time $\tau_z$ as

\begin{equation}
    \tau'_z = k\tau_z
\end{equation}

\noindent
where $k$ is a constant. The larger the constant, therefore, means the weaker the radiation damping. Chromaticity 5 was chosen for this study due to its high growth rate, increasing the visibility of the TSBI.

\begin{figure}
    \centering
    \includegraphics[width=\linewidth]{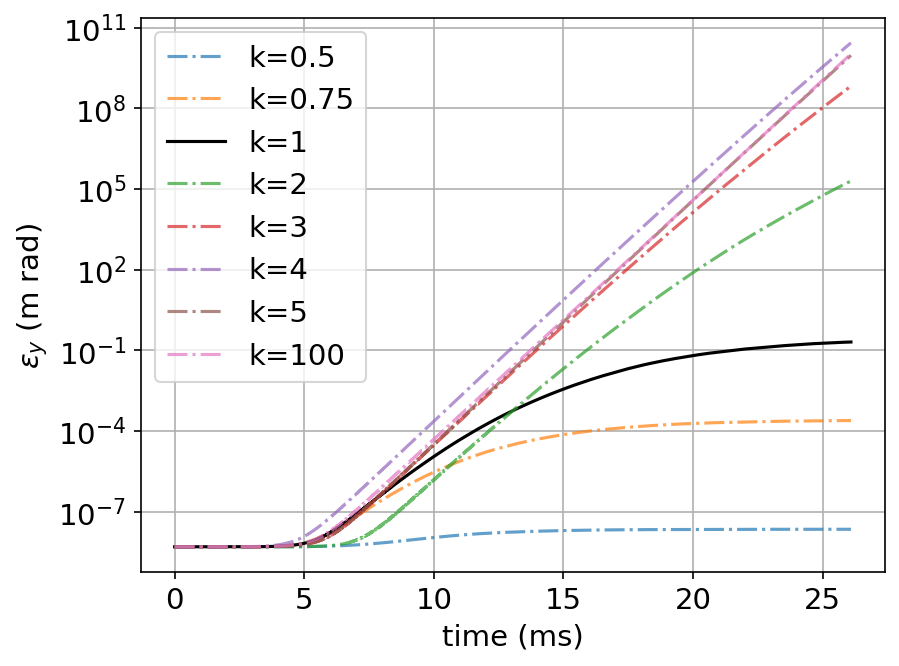}
    \caption{Vertical emittance growth influenced by different longitudinal synchrotron damping times without ADTS. Parameters: $E_0=1.9$\,GeV (at 80\,ms on the ramp), $\xi=5$, $q=1.5$\,nC.}
    \label{fig:long_damping_effect}
\end{figure}

The results on the vertical emittance evolution during some twenty milliseconds at different $k$ at a fixed energy 1.9\,GeV are shown in Fig.\,\ref{fig:long_damping_effect}. The case with $k=1$ is where the nominal damping time similar to all previous studies was used, and that with $k=100$ corresponds to a negligible longitudinal damping. An astonishing result can be seen clearly for $0.5 \leq k \leq 1$; that is, the emittance growth slows down over time as if something was counteracting the instability. Obviously, this effect must come from the damping in the longitudinal plane since the emittance saturation is reached sooner as $k$ is reduced, and vice versa.

From Ref.\,\cite{Suzuki:167026}, it is shown that an approximate stability criterion for the transverse coupled-bunch instability is that 

\begin{equation}
    \text{(growth rate)} < \frac{1}{\tau_\beta} + \frac{\abs{m}}{\tau_z}
\end{equation}

\noindent
where $\tau_\beta$ is the transverse damping time and $m$ is the azimuthal head-tail mode number. It can then be deduced that the right-hand side of the inequality is nothing more than the total damping time in the transverse plane. This implies that longitudinal damping can really modify the damping time in the transverse plane when $\abs{m}\geq1$.

The results in Fig.\,\ref{fig:long_damping_effect} agree with the study in \cite{lindberg:PhysRevAccelBeams.24.024402} where the effect of longitudinal damping on TCBI was investigated. In this reference, it is also shown that this effect is stronger when the longitudinal damping rate is larger. However, in our case, when we look at different chromaticities at the nominal longitudinal damping rate, the results do not follow the same conclusion as in this reference. We found that chromaticity 5 is the least affected by the longitudinal damping, and the lower the chromaticity, the stronger the effect. Meanwhile, in the reference, the effect of longitudinal damping is more important at higher chromaticity.

\begin{figure}
    \centering
    \includegraphics[width=\linewidth]{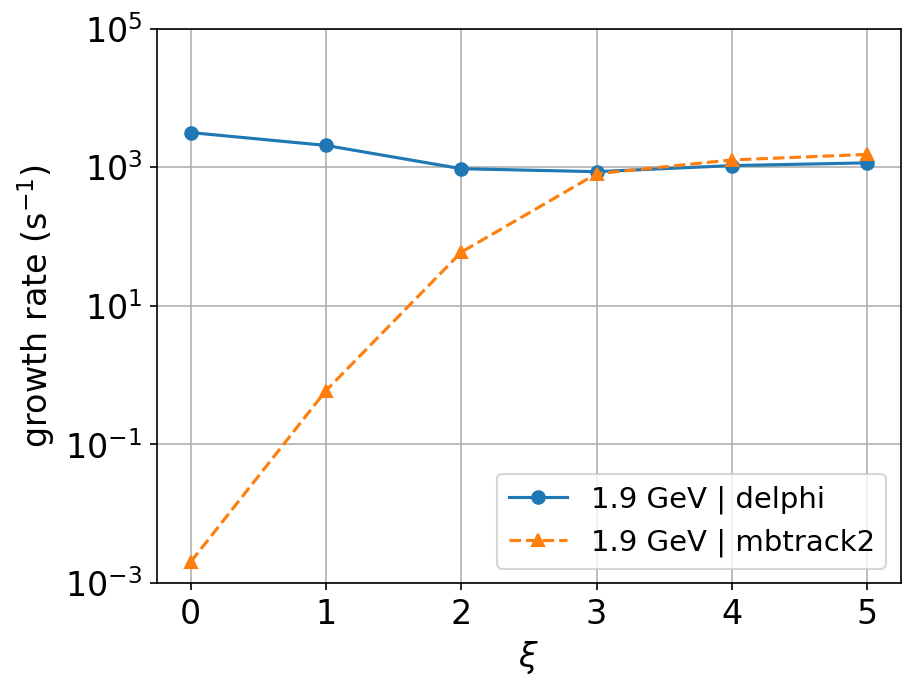}
    \caption{Comparison of the TSBI growth rate obtained from \texttt{mbtrack2} and \texttt{DELPHI} at 1.9\,GeV and 1.5\,nC.}
    \label{fig:benchmark_delphi_1_9GeV}
\end{figure}

A benchmark on the growth rate as a function of chromaticity was done with a Vlasov solver \texttt{DELPHI} \cite{delphi} and the result is shown in Fig.\,\ref{fig:benchmark_delphi_1_9GeV}. Both codes were given with the same parameters and the same bunch length at each energy. The only difference is that \texttt{DELPHI} does not have radiation damping in any plane, while \texttt{mbtrack2} still has one in the longitudinal plane. It evidently shows that the longitudinal radiation damping plays a role only up to chromaticity 2.

The reason for the disagreement between our result and the one in \cite{lindberg:PhysRevAccelBeams.24.024402} possibly lies in the impedance used in this study. It is composed of a resistive-wall impedance and a large bump from a broad-band resonator (see Fig.\,\ref{fig:Zyeff_chro1}). Such an impedance composition can result in the \textit{head-tail damping} effect, which we shall fully discuss in Section \ref{sec:headtail_damping}. In brief, this effect is important at low chromaticity when the beam spectrum can see the stabilizing effect coming from the broad-band resonator. However, at larger chromaticity, this favorable effect is lost since the beam spectrum is shifted far away from the broad-band resonator bump. This mechanism could diminish the damping effect diffused from the longitudinal plane due to the high growth rate.
 
In any case, it has been shown that longitudinal damping can seriously affect transverse instabilities by altering the transverse damping time. Therefore, longitudinal synchrotron radiation should always be included in tracking simulations. 

\section{Transverse Coupled-bunched Instability} \label{sec:MB}

\subsection{General trend from analytical estimation}\label{sec:general_trend_TCBI}

One of the most important changes from a 3GLS booster to a 4GLS one might be the reduction in vacuum chamber radius. It is one way to increase the magnetic flux density on the beam by letting the magnetic poles approach the beam as much as possible. However, the smaller vacuum chamber size inevitably raises concerns over the TCBI driven by RW impedance, also referred to as Resistive-Wall Instability (RWI). Since its growth rate is inversely proportional to the vacuum chamber radius cubed, at least at zero chromaticity $\xi=0$, as shown in the following equation \cite{Nagaoka:xe5009}:

\begin{equation}\label{eq:RWI_growth_rate}
    \qty(\tau^{-1}_\text{RWI})_{\xi=0} = \frac{\beta_0\omega_0I}{4\pi E_0} \frac{R}{b^3} \qty[\frac{2cZ_0}{(1-\nu_\beta)\sigma\omega_0}]^{1/2}
\end{equation}

\noindent
where $\beta_0$ is the average optic beta function, $\omega_0$ is the angular revolution frequency, $I$ is the total beam current, $E_0$ is the beam energy in eV, $R$ is the machine radius, $b$ is the chamber radius, $c$ is the speed of light, $Z_0$ is the impedance of free space, $\nu_\beta$ is the fractional part of the betatron tune, and $\sigma$ is the conductivity of the chamber wall. Apart from the vacuum chamber radius dependence, we can also see that the RWI growth rate is inversely proportional to the beam energy, $\tau_\text{RWI}^{-1}\propto E_0^{-1}$. It should be mentioned that Eq. (\ref{eq:RWI_growth_rate}) was proved from a uniform beam filling pattern. Thus, in a case where the ring is partially filled, as ours, it can only give an estimated growth rate.

We can say that the natural countermeasure to most collective instabilities for electron machines is synchrotron radiation damping, as we have just seen from the sawtooth instability investigation. It is, however, effective only at high energy since its damping rate is given by

\begin{equation}\label{eq:rad_damping_rate}
    \tau_\text{rad}^{-1} = \frac{j}{2} \frac{U_0}{E_0T_0}
\end{equation}

\noindent
where $j$ is the damping partition number of the interested plane, $U_0$ is the energy loss per turn, and $T_0$ is the beam revolution period. Knowing that $U_0\propto E_0^4$ \cite{Wolski:CAS2014}, we can deduce that $\tau_\text{rad}^{-1}\propto E_0^{3}$.

\begin{figure}[tb]
    \centering
    \includegraphics[width=\linewidth]{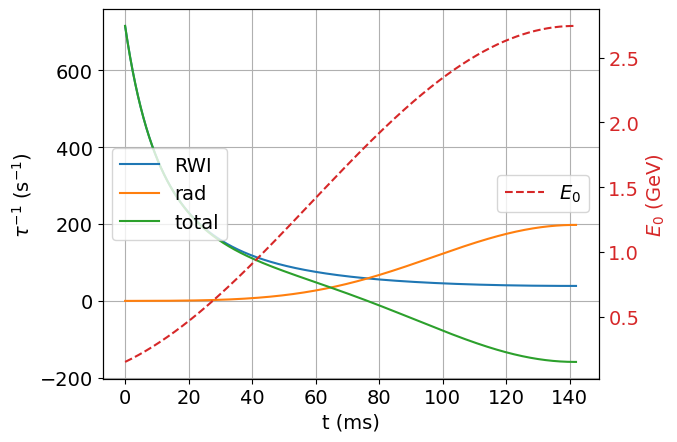}
    \caption{The zero-chromaticity RWI growth rate, radiation damping rate, total growth rate, and beam energy along the SOLEIL\,II booster's ramp.}
    \label{fig:gr_compare}
\end{figure}

The opposite direction of energy dependence of $\tau_\text{RWI}^{-1}$ and $\tau_\text{rad}^{-1}$ makes RWI and synchrotron radiation rivals along the ramp. Figure \ref{fig:gr_compare} shows the evolution of these two parameters along the ramp in the case of the SOLEIL\,II booster. The \textit{total growth rate} which is defined as

\begin{equation}
    \tau^{-1} =  \tau_\text{RWI}^{-1} - \tau_\text{rad}^{-1}
\end{equation}

\noindent
is also shown. Note that, here, we are interested in the zero chromaticity circumstance, which can only provoke the mode $m=0$. Theoretically, the longitudinal damping effect can then be neglected. A positive total growth rate means that instability is stronger than synchrotron radiation, and the beam center-of-mass (CM) oscillation can grow exponentially in the transverse plane, which is the sign of RWI. The point where the total growth rate goes into the negative side, therefore, marks the point where synchrotron radiation is strong enough to overtake RWI and possibly stabilize the beam. It can be seen in this figure that synchrotron radiation is ineffective during the first half of the ramp, resulting in possible instability.

\begin{figure}[tb]
    \centering
    \includegraphics[width=\linewidth]{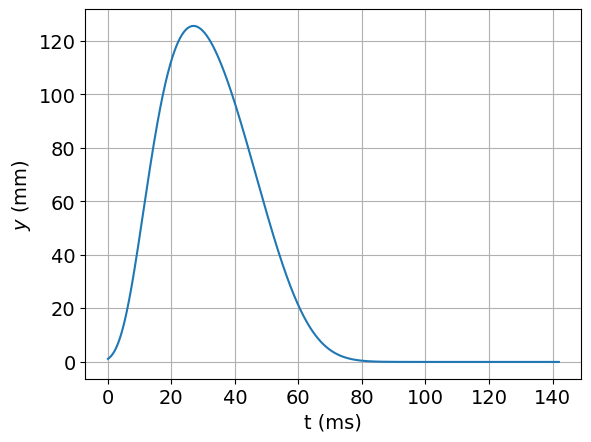}
    \caption{Predicted CM oscillation amplitude along the SOLEIL\,II booster ramp according to Eq. (\ref{eq:yt_analytical}). Initial position $y_0=\SI{1}{mm}$.}
    \label{fig:yt_analytical}
\end{figure}

To quickly predict the beam behavior, we define a simple equation to describe the beam CM transverse oscillation amplitude $y$ as a function of time $t$:

\begin{equation}
    y(t) = y_0e^{t/\tau(t)}, \label{eq:yt_analytical}
\end{equation}

\noindent
with $y_0$ being the initial amplitude and $1/\tau=\tau^{-1}$ being the total growth rate defined previously. This equation represents the CM as a point particle whose amplitude can only either grow or damp exponentially depending on the sign of $\tau^{-1}$. Figure\,\ref{fig:yt_analytical} visualizes Eq. (\ref{eq:yt_analytical}) when $\tau^{-1}$ is that of the SOLEIL\,II booster and $y_0$ is deliberately set at 1\,mm. The result shows that an enormous CM oscillation of more than 120 times $y_0$ can be expected right after the injection due to the strong RWI and weak synchrotron radiation. The oscillation is then damped at higher energy when the synchrotron radiation damping rate overtakes the RWI growth rate.

This result gives a general trend of RWI in every booster at zero chromaticity when there are no other stabilizing mechanisms involved, such as Landau damping, head-tail damping, or a transverse feedback system. The height of the curve depends on each machine's ramp speed: the faster the ramp, the sooner synchrotron radiation can suppress RWI, and the smaller the peak oscillation will be. Another important factor is the vacuum chamber radius. For 3GLS boosters whose chamber radius is typically large, the RWI growth rate is small, so the CM oscillation is relatively small as a consequence.

\subsection{Simulation results}
Tracking simulations were also done by the code \texttt{mbtrack2} and the single-bunch and multibunch collective effects were calculated simultaneously. The code implementation is largely inspired by the code \texttt{mbtrack} \cite{Skripka:2016221}. It should be noted that the single-bunch tracking is indispensable even if we are interested in a multibunch effect. This is because the former takes into account the self-field interaction which will affect the bunch longitudinal dynamics and, therefore, affect the wake potential calculation that depends on the bunch longitudinal profile.

The simulations were carried out at the nominal multi-bunch charge 3\,nC (equivalent to 5.7\,mA) and at the maximum technologically achievable charge for the SOLEIL\,II booster of 10\,nC (19.2\,mA). These charges are distributed over 104 bunches of the total length of 295\,ns, corresponding to the injection scheme for the 416-bunch operation mode of the storage ring. Knowing that the booster harmonic number is 184, this filling pattern leaves 80 empty buckets after the bunch train. 

The cavity voltage was modeled as a sinusoidal wave without taking into account the transient beam loading effect since the beam current is small, even at the maximum charge.

\begin{figure}[tb]
    \centering
    \begin{subfigure}[b]{0.49\textwidth}
        \includegraphics[width=0.9\textwidth]{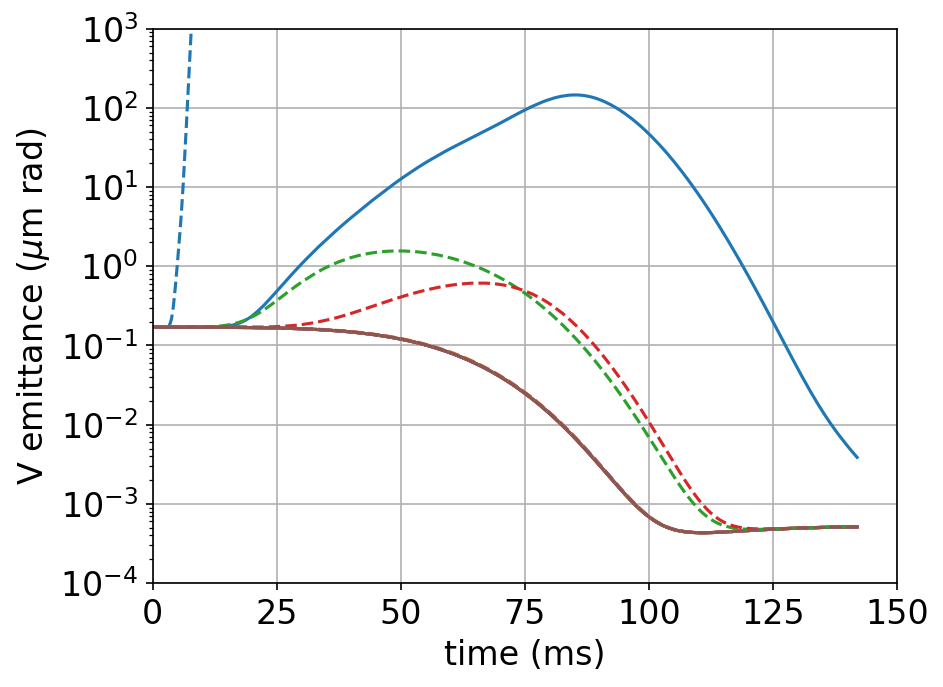}
        \caption{Without ADTS}
        \label{fig:MB_ramp_noADTS}
    \end{subfigure}
    \begin{subfigure}[b]{0.49\textwidth}
        \includegraphics[width=0.9\textwidth]{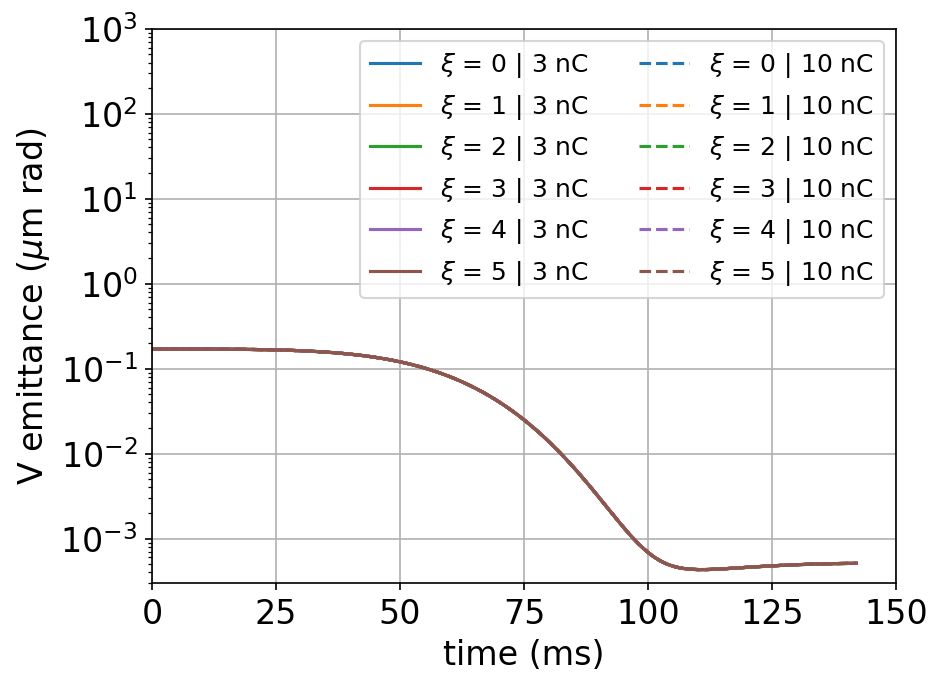}
        \caption{With ADTS}
        \label{fig:MB_ramp_ADTS}
    \end{subfigure}
    \caption{Average vertical beam emittance along the ramp at different chromaticities and charges. Both sub-figures share the same legend. All curves in (b) are identical.}
    \label{fig:MB_ramp}
\end{figure}

Figure\,\ref{fig:MB_ramp} shows the simulated vertical emittance averaged over all 108 bunches in the beam. In Fig.\,\ref{fig:MB_ramp_noADTS}, the ADTS was excluded from the tracking, leaving the beam with only synchrotron radiation to counteract the instability. It is evident that the beam at both charges, especially at low chromaticities, can have an emittance blow-up at low energy before being damped at high energy if we do not have any or not strong enough nonlinearity, thus no Landau damping. This result agrees with the previously predicted general trend of TCBI in a booster.

In contrast, in the case where ADTS is included as shown in Fig.\,\ref{fig:MB_ramp_ADTS}, the beam does not suffer from instability, whether at 3 or 10\,nC and regardless of chromaticity. It can then be inferred that the beam is also stable for other charges below 10\,nC. This favorable outcome is due to the large tune spread induced by the ADTS \cite{Foosang_2024}. It emphasizes the beneficial side of the nonlinearity that comes with a 4GLS booster. At the same time, it reminds us that adding nonlinearity in tracking simulation for such a machine is important due to the large impact it can have on the results.

\subsection{Head-tail damping}\label{sec:headtail_damping}

\begin{figure}[tb]
    \centering
    \includegraphics[width=\linewidth]{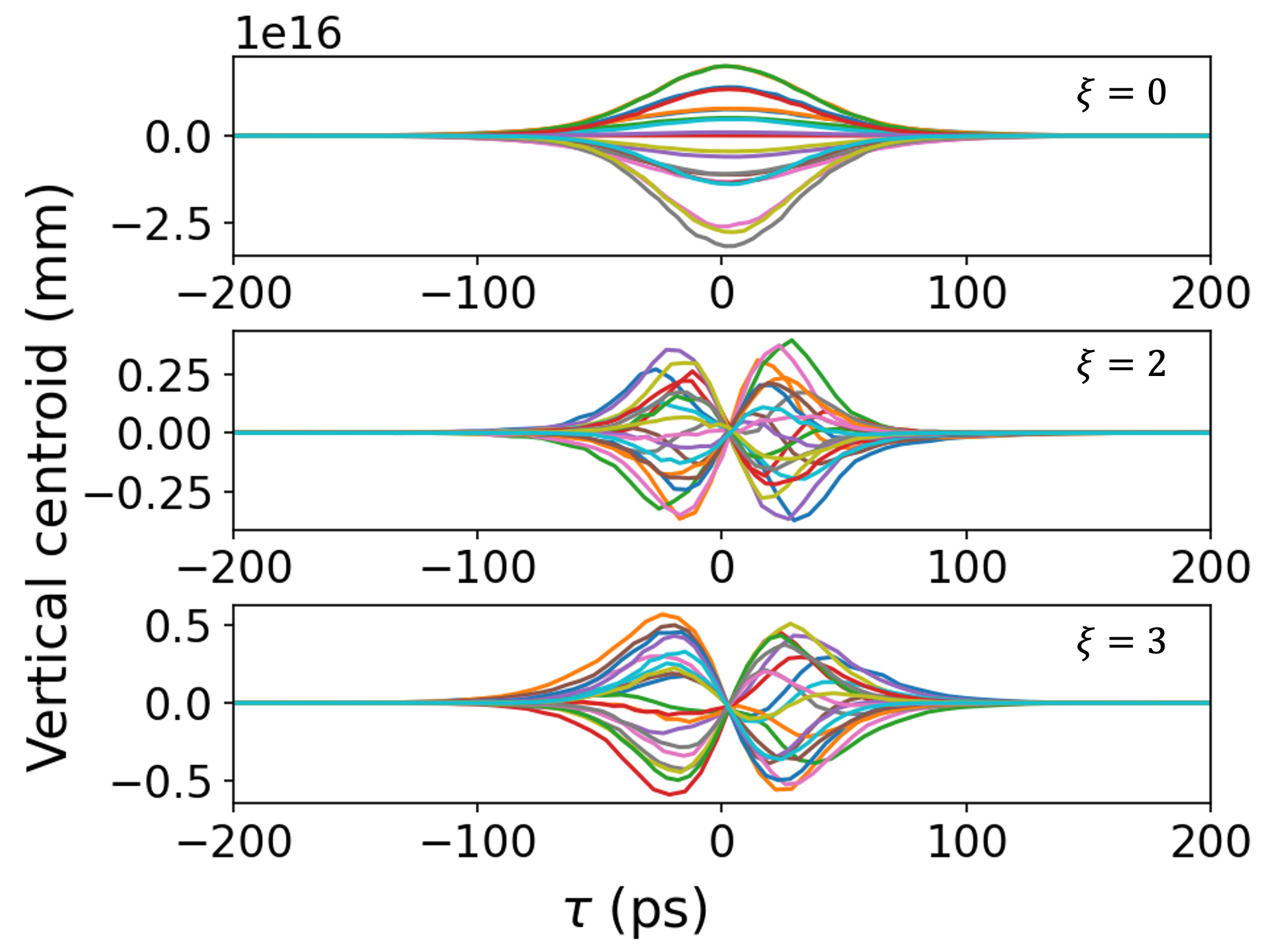}
    \caption{Vertical centroid position of a bunch around 50-60 ms of the 10 nC beam at chromaticity 0-3 without ADTS. The horizontal axis is the bunch longitudinal position in time relative to the rf bucket center.}
    \label{fig:Vcentroid_MB_10nC}
\end{figure}

Now, we will demonstrate the importance of broad-band impedance in stabilizing coupled-bunch instability. Let us consider the case where ADTS are excluded so that their effect does not interfere with our analysis. If we pay close attention to the 10\,nC cases in Fig.\,\ref{fig:MB_ramp_noADTS} (dashed curves), we will see that the beam is unstable at zero chromaticity, then perfectly stable at chromaticity 1, before going unstable again at chromaticity 2 and 3. Generally, one would expect the growth rate to drop proportionally to the chromaticity when the latter is small. This jump is, therefore, relatively surprising since it shows some kind of discontinuity in the instability growth rate with chromaticity. Often, such behavior is a sign of switching from one head-tail mode to another higher one, but Fig. \ref{fig:Vcentroid_MB_10nC} shows us that the observed head-tail mode $m$ remains indeed $m=1$ due to only one node being seen from the bunch centroid position, which is a product between the bunch dipole moment and the longitudinal bunch profile. 

% \begin{figure}[tb]
%     \centering
%     \begin{subfigure}[b]{0.49\textwidth}
%         \includegraphics[width=0.9\textwidth]{fig11a.png}
%         \caption{$\xi=2$}
%         \label{fig:Vcentroid_MB_chro2_10nC}
%     \end{subfigure}
%     \begin{subfigure}[b]{0.49\textwidth}
%         \includegraphics[width=0.9\textwidth]{fig11b.png}
%         \caption{$\xi=3$}
%         \label{fig:Vcentroid_MB_chro3_10nC}
%     \end{subfigure}
%     \caption{Vertical centroid position of a bunch around 50-60 ms of the 10 nC beam at chromaticity 2 and 3 without ADTS. The horizontal axis is the bunch longitudinal position in time relative to the rf bucket center.}
%     \label{fig:Vcentroid_MB_10nC}
% \end{figure}

The explanation does involve the single-bunch scale mechanism usually called the \textit{head-tail damping}. It refers to the situation where the TCBI is suppressed by shifting the chromaticity to a positive value. It does not compete with instability with its own damping rate like, for example, the synchrotron radiation, but rather directly reduces the instability growth rate. This is because the theoretical TCBI growth rate is proportional to \textit{effective impedance} \cite{Pedersen:254071}, which is a sum of the transverse impedance weighted by the bunch power spectrum resulting from the head-tail oscillation. In other words, it is the net quantity of the impedance that can be sampled by the bunch frequency. One can see it as an overlapping area between the impedance and the bunch spectrum. The shape of the impedance, the head-tail oscillation mode, and the chromaticity can have an impact on how these two variables overlap.

Also, in \cite{Pedersen:254071}, it is explained that a broad-band impedance appearing at high frequency in addition to a resistive-wall one can yield the head-tail damping effect on TCBI. This is because the broad-band part gives rise to the impedance on the positive frequency side, which contributes to a negative TCBI growth rate. When the bunch spectrum is also shifted towards the positive frequency due to positive chromaticity, it can then see the broad-band impedance and counterbalance the impedance sampled on the negative frequency side that contributes to a positive growth rate. However, the head-tail damping effect can be broken when the chromaticity keeps increasing since the bunch spectrum will be shifted away from the hill of the broad-band impedance.

\begin{figure}[tb]
    \centering
    \includegraphics[width=\linewidth]{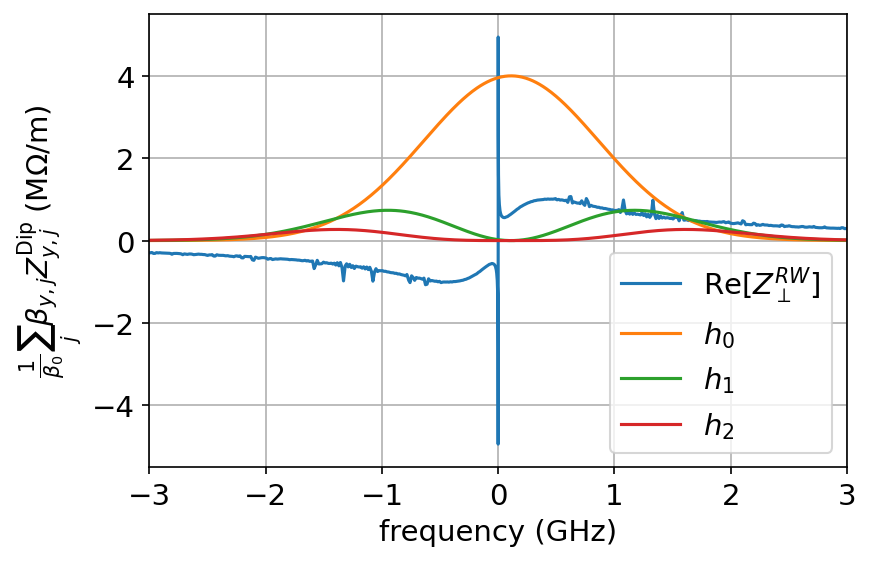}
    \caption{The real part of the total vertical dipolar impedance, weighted by beta function, of the SOLEIL\,II booster plotted with the head-tail oscillation spectrum $h$ of mode $m=0-2$ (arbitrary unit) whose center frequency is shifted by 111.9\,MHz due to chromaticity 1.}
    \label{fig:Zyeff_chro1}
\end{figure}

In case of the SOLEIL\,II booster, its vertical dipolar impedance is shown in Fig.\,\ref{fig:Zyeff_chro1} and Table \ref{tab:growth_rate} summarizes the effective impedance and the theoretical TCBI growth rate at the injection energy 150\,MeV at chromaticities 1--3 for the first three head-tail modes (see the equations used for the calculation in Appendix \ref{app:eff_imp}). For chromaticities 2 and 3, as we have deduced from Fig.\,\ref{fig:Vcentroid_MB_10nC}, the dominant mode that represents the instability is $m=1$, in agreement with this calculation. For chromaticity 1, however, we cannot know for sure which mode dominates the beam since the instability does not occur. In any case, if we pick its strongest mode $m=1$ and compare it with the same mode of the other chromaticities, we can see that the growth rate at chromaticity 1 is indeed smaller than that at chromaticity 2 and 3.

\begin{table}[tb]
    \centering
    \footnotesize
    \caption{Vertical effective impedance and the theoretical TCBI growth rate at 150\,MeV (Parameters: $n=103$, $q=10$\,nC, $M=104$, $\sigma_\tau=150$\,ps)}
    \begin{ruledtabular}
    \begin{tabular}{cccc}
        $\xi$   &   $m$ & $Z_{y,\text{eff}}$ (k$\Omega$) & growth rate (s$^{-1}$) \\ 
    \midrule
        1       &   0   &   $160 + 1058i$       &   -3419   \\
                &   \textbf{1}   &   $\mathbf{-41 + 398i}$        &   \textbf{437}     \\
                &   2   &   $-46 + 231i$        &   328     \\
    \midrule
        2       &   0   &   $245 + 1037i$       &   -5233   \\
                &   \textbf{1}   &   $\mathbf{-66 + 4245i}$       &   \textbf{704}     \\
                &   2   &   $-77 + 242i$        &   551     \\
    \midrule
        3       &   0   &   $323 + 1002i$       &   -6901   \\
                &   \textbf{1}   &   $\mathbf{-78 + 466i}$        &   \textbf{834}     \\
                &   2   &   $-106 + 261i$       &   755     \\
    \end{tabular}
    \end{ruledtabular}
    \label{tab:growth_rate}
\end{table}

This is effectively due to the shape of the impedance. In a 4GLS machine, one might expect the resistive-wall impedance to dominate since the vacuum chamber radius tends to be small and use only this to estimate the machine impedance. But, Fig.\,\ref{fig:Zyeff_chro1} has shown us that our booster is actually composed of two main parts; a resistive-wall impedance which peaks at zero frequency and a broad-band impedance which peaks around 450--500\,MHz. The first part, of course, comes from the vacuum chamber around the machine. The second part appears due to the presence of three ceramic chambers whose inner wall is coated with a 200\,nm thick titanium layer, used to accommodate the fast kicker magnets at the injection and extraction points \cite{foosang:tel-04496043}. 

\begin{figure}
    \centering
    \includegraphics[width=\linewidth]{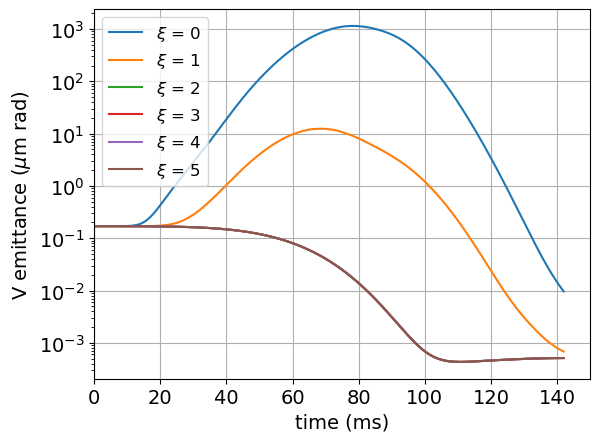}
    \caption{Average vertical beam emittance along the ramp at 3\,nC without ADTS and without broad-band impedance.}
    \label{fig:TCBI_noBB_noADTS_3nC}
\end{figure}

Such an impedance shape damps the head-tail mode $m=0$, whose power amplitude is relatively large, and gives rise to other higher-order modes with a much lower amplitude but have a better overlap with the impedance. Without the broad-band part in the impedance model, the effect of head-tail damping will be much weaker so that it cannot stabilize the beam. This is what we can observe in Fig.\,\ref{fig:TCBI_noBB_noADTS_3nC} which shows the emittance along the ramp of a 3\,nC beam without ADTS and without the broad-band impedance, only the resistive-wall one. When compared to Fig.\,\ref{fig:MB_ramp_noADTS}, we can notice a larger emittance at zero chromaticity and that the beam at chromaticity 1 is not anymore stable.

In summary, the results in this section tell us that broad-band impedance can be beneficial for TCBI since it can reduce the effective impedance by damping the powerful head-tail mode 0 and exciting weaker higher-order modes. The particular shape of the higher-mode bunch spectrum can then result in a nonlinear relationship between the chromaticity and TCBI growth rate. This, however, does not mean that broad-band resonators should be imprudently introduced in the machine since they can worsen the instabilities in the single-bunch regime, such as increasing TMCI and head-tail growth rate or reducing microwave instability threshold. Instead, it tells us that care should be taken when one wants to model a machine impedance for a TCBI study, not to neglect possible broad-band sources.

\section{Conclusion}
Due to the foreseen high impedance and nonlinearity in 4GLS booster synchrotrons, it requires that collective effects in such machines are examined more profoundly than what was usually done for the 3GLS ones. Transverse beam instabilities along the ramp in a low-emittance booster, therefore, have been studied both in the single-bunch and multibunch scales, using the SOLEIL\,II booster as the case study.

In the single-bunch regime, the beam can be unstable at the injection energy, but the stronger synchrotron radiation damping along the ramp will help suppress the instabilities. However, in case of a high charge, the growth rate can be exceedingly high, resulting in an emittance blow-up immediately after the injection. In this scenario, ADTS is an efficient mechanism that can limit the growth due to the strong Landau damping induced by the large emittance. Nevertheless, if the growth rate stays higher than the synchrotron damping rate, the sawtooth instability can appear at high energy since the ADTS becomes less efficient due to the damped emittance.

It has been confirmed that transverse instabilities can be suppressed by longitudinal damping. The synchrotron radiation in the longitudinal plane substantially helped suppress the emittance growth in the transverse plane. This effect, however, only appears at high energy when synchrotron radiation is strong to a certain degree. This result illustrates that longitudinal damping can seriously impact the transverse beam dynamics and must not be excluded in tracking simulations.

For the multibunch regime, the studies have shown that ADTS can be an effective mechanism against TCBI as it can sustain the beam stability throughout the ramp in the SOLEIL\,II case because of Landau damping similar to the single-bunch case.

Lastly, we have confirmed that a broad-band impedance at high frequency can result in a suppression of TCBI at slightly positive chromaticity due to the head-tail damping effect, as explained in \cite{Pedersen:254071}. 

To conclude, in addition to synchrotron radiation, whose effect is well known, we can deduce from this work that energy ramp, ADTS, longitudinal damping, and broad-band impedance are four necessary components to describe accurately TSBI and TCBI in boosters. Therefore, to design an ultra-low emittance booster synchrotron, consideration of these effects is imperative.

\section*{Acknowledgements}
    Parts of this work were done with the support of the Franco-Thai Scholarship, financed by the French Embassy in Thailand, received during the first author's PhD program. The simulations in the multibunch regime presented in this article were made possible by the CCRT HPC resource (TOPAZE supercomputer) hosted at Bruy\`eres-le-Ch\^atel, France. The authors would also like to thank M.-A. Tordeux and P. Alexandre for the valuable discussion and effective collaboration on the SOLEIL\,II booster project.

\appendix

\section{ADTS}\label{app:adts}

\begin{figure}
    \centering
    \includegraphics[width=\linewidth]{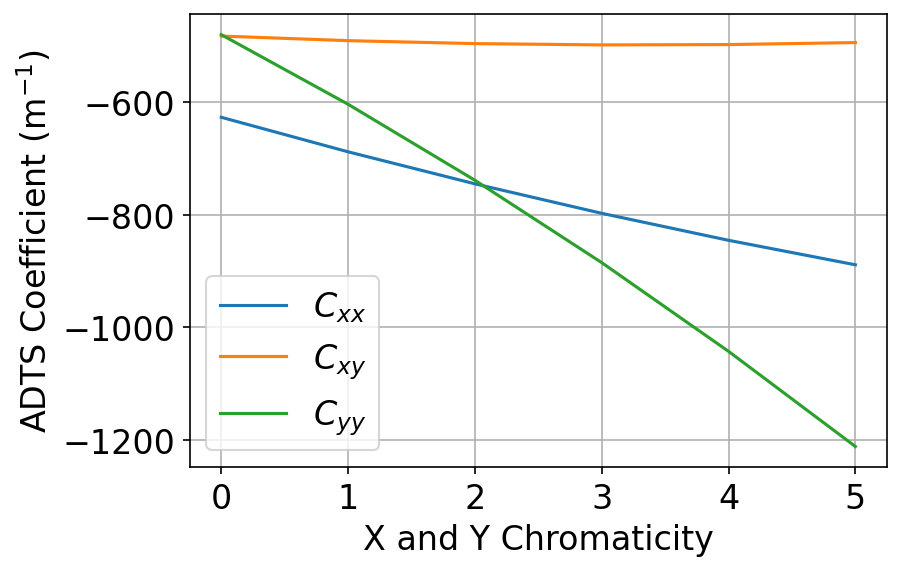}
    \caption{SOLEIL\,II booster's theoretical ADTS coefficients.}
    \label{fig:adts_coef}
\end{figure}

The Amplitude-Dependent Tune Shift (ADTS) of a single particle implemented in \texttt{mbtrack2} and used in this study is described by the following so-called second-order equation: 

\begin{equation}
    \begin{pmatrix}
        \Delta\nu_x \\ 
        \Delta\nu_y
    \end{pmatrix}
    = 
    \begin{pmatrix}
        C_{xx} & C_{xy} \\
        C_{yx} & C_{yy}
    \end{pmatrix}
    \begin{pmatrix}
        J_x \\
        J_y
    \end{pmatrix},
\end{equation}

\noindent
where $\Delta\nu_{i}$ is the tune shift, $C_{ij}$ is an ADTS coefficient, and $J_{i}$ is the Courant-Snyder invariant of the particle. The indices $i$ and $j$ stand for $x$ if the plane considered is the horizontal one or $y$ if it is the vertical one. The theoretical ADTS coefficients of the SOLEIL\,II booster at different chromaticities obtained from the Accelerator Toolbox (AT) code \cite{pyAT} are shown in Fig.~\ref{fig:adts_coef}. The horizontal and vertical chromaticities are identical at each point. Here, the cross-terms in the coefficient matrix are shown as one as their values are identical, i.e. $C_{xy}=C_{yx}$.

\section{TCBI growth rate and effective impedance}\label{app:eff_imp}
The theoretical TCBI growth rate in general can be given by \cite{Pedersen:254071}

\begin{equation}\label{eq:TCBI_tuneshift}
    \tau_\text{TCBI}^{-1} = -\Im[\Delta f_{m,n}] = \frac{-1}{m+1} \frac{e\beta I_b}{8\pi^2\nu_\beta f_0\gamma m_0 \tau_L} \Re[Z_{\perp,\text{eff}}]
\end{equation}

\noindent
where $\Delta f_{m,n}$ is the complex frequency shift, $m$ is the head-tail mode number, $e$ is the elementary charge, $\beta$ and $\gamma$ are the usual relativistic factors, $I_b$ is the current per bunch, $\nu_\beta$ is the betatron tune, $f_0$ is the revolution frequency, $m_0$ is the particle's rest mass, $\tau_L$ is the bunch length in time. The last parameter $Z_{\perp,\text{eff}}$ is the effective impedance: \cite{Pedersen:254071, Sacherer:bunched-beam-paper}

\begin{equation}\label{eq:effective_impedance_def}
    Z_{\perp,\text{eff}} = \frac{\sum_{p=-\infty}^{\infty} Z_\perp(f_p)h_m(f_p-f_\xi)}{\sum_{p=-\infty}^{\infty}h_m(f_p-f_\xi)}.
\end{equation}

\noindent
where $h_m$ is the power spectrum of the head-tail mode $m$, $f_p$ is the coupled-bunch mode frequencies given by \cite{Pedersen:254071}

\begin{equation}
    f_p = (n+pM+\nu_\beta)f_0 + mf_s
\end{equation}

\noindent
and $f_\xi$ is the frequency shift of the bunch spectrum due to chromaticity defined as \cite{Pedersen:254071}

\begin{equation}
    f_\xi=\frac{\xi}{\eta}\nu_\beta f_0
\end{equation}

\noindent
Here, $n$ is the coupled-bunch mode number, $p$ is an integer that can be both negative and positive, $M$ is the number of bunches, $f_s$ is the synchrotron oscillation frequency, and $\eta$ is the phase slip factor. It is this $h_m$ that links the single-bunch mechanism to the multibunch one. For a Gaussian bunch, the head-tail power spectrum takes the form \cite{chao:handbook}

\begin{equation}
    h_m(f) = \frac{1}{2^m m!}(2\pi f\sigma_\tau)^{2m} e^{-(2\pi f \sigma_\tau)^2}
\end{equation}

\noindent
where $\sigma_\tau$ is the rms bunch length.

\bibliography{biblio}

\end{document}